\documentclass{rspublic}

\begin{document}

\title[Starbursts, Galaxy Formation, \& AGN]{The Role of Starbursts in the
Formation of Galaxies \& Active Galactic Nuclei}

\author[T. Heckman]{Timothy M. Heckman}

\affiliation{Department of Physics \& Astronomy, Johns Hopkins University,
Baltimore, MD 21028 USA}

\label{firstpage}

\maketitle

\begin{abstract}{Starbursts, Galactic Winds, Seyfert Galaxies, Quasars}
Starbursts are episodes of intense star-formation in the
central regions of galaxies, and are the sites of roughly 25\% of the high-mass
star-formation in the local universe.
In this contribution I review the role starbursts play in the
formation and evolution of galaxies, the intergalactic medium, and active
galactic nuclei.
First, I point out the empirical similarities between local starbursts
and the Lyman-break population at high redshift, and emphasize the
implied similarities in their basic physical, dynamical, and chemical
properties. In the local universe, more massive galaxies host
more luminous, more metal-rich, and dustier (IR-dominated) starbursts.
This underscores
the need for a panchromatic approach to documenting and understanding
the cosmic history of star-formation. Second, I review the 
systematic properties of starburst-driven galactic superwinds. These drive
metal-rich {\it dusty} gas outward
at a typical velocity
of 400 to
800 km s$^{-1}$ (independent of the galaxy rotation speed) and at 
several times the star-formation rate. They can be directly observed
both in local starbursts and high-redshift galaxies. They are probably
responsible for establishing the strong mass-metallicity relation in spheroids
and for the metal-enrichment and (pre)heating of the inter-galactic medium.
They {\it may} have also ejected cosmologically-significant amounts
of intergalactic dust.
Third, I discuss UV observations of the nuclei
of type 2 Seyfert galaxies. These show that compact (few-hundred-pc-scale)
heavily-reddened starbursts are the source of most of the `featureless
continuum' in UV-bright Seyfert 2 nuclei, and are an energetically
significant component in these objects. Finally, I discuss
the evolution of the host galaxies of radio-quiet
quasars. Rest-frame optical images imply that the hosts
at $z \sim$ 2 are
only as luminous as present-day $L_*$ galaxies, less massive than
than the hosts of similarly-luminous low-z quasars, similar to
the Lyman-break galaxies, and much less luminous than powerful
radio galaxies at the same redshift. These results are consistent
with the idea of hierarchical galaxy assembly, and suggest that
supermassive black holes may be formed/fed before their host galaxy is
fully assembled.
\end{abstract}

\section{Introduction}

Starbursts are short-lived episodes of intense star-formation that usually
occur in the
`circum-nuclear' (kpc-scale) regions of
galaxies, and dominate the integrated emission from the `host'
galaxy (Leitherer \textit{et al.} 1991).
Starbursts are major components of the local universe (e.g.
Gallego \textit{et al.} 1995; Soifer \textit{et al.} 1987), and are the sites of
$\sim$ 25\% of the total (high-mass) star-formation in the local universe
(Heckman 1997).

The cosmological relevance of starbursts has been dramatically
underscored by the spectacular discovery of
populations of high-redshift ($z >$ 2) star-forming
field galaxies selected by their rest-frame ultra-violet continuum emission
(Steidel \textit{et al.} 1999; Lowenthal \textit{et al.} 1997), rest-frame
far-infrared emission (e.g. Hughes \textit{et al.} 1998;
Barger, Cowie, \& Richards 1999; Cowie, this conference),
Ly$\alpha$ emission (Hu, Cowie, \& McMahon 1998), and H$\alpha$ emission
(Teplitz, Malkan, \& McLean 1998; Mannucci \textit{et al.} 1998).
The co-moving space
density and luminosity of these galaxies
imply that they
almost certainly represent precursors of typical present-day galaxies
and are responsible for the production of a significant fraction of the
stars and heavy elements in the present-day universe (e.g. Madau
\textit{et al.} 1996;
Blain \textit{et al.} 1998; Calzetti \& Heckman 1998).

Starbursts may also play a vital energetic or evolutionary role in
active galactic nuclei (AGN). 
There have been recurring suggestions to this effect in the Seyfert galaxy
phenomenon (e.g. Weedman 1983; Perry \& Dyson 1985; Terlevich \&
Melnick 1985; Norman \& Scoville 1988; Cid Fernandez \& Terlevich
1995). On a more global scale, the rough proportionality between
the mass of a supermassive black hole and that of the stellar spheroid
in which it now resides (Magorrian \textit{et al.} 1998; van der Marel 1999)
strongly suggests that the quasar phenomenon
is an intimate part of the formation or early evolution of massive
ellipticals and bulges. 

In this contribution I will therefore discuss the relevance of local starbursts 
to understanding the high-redshift universe and its major baryonic components:
star-forming galaxies, AGN, and the inter-galactic medium.
In \S 2 I will argue that starbursts are the only local analogues to
the star-forming galaxies observed at high-redshift, and will summarize
some of inferences that follow. In \S 3 I will describe the systematic
properties of starburst-driven galactic superwinds and summarize the
cosmological implications of these outflows. In \S 4 I will present a
status report on efforts to understand the energetic/evolutionary
significance of young stars in/near the nuclei of Seyfert galaxies. Then,
in \S 5 I will report new results on the properties of the host
galaxies of high-redshift radio-quiet quasars, and compare these to
theoretical expectations.

\section{Starbursts as Analogues to High-z Galaxies}

Starburst galaxies are the only plausible local analogues to the population
of star-forming galaxies at high-redshift. Meurer \textit{et al.} (1997)
showed that local starbursts and high-redshift Lyman-break galaxies have
similar rest-frame UV surface brightnesses and UV colors.
Using the empirical correlation between UV color and extinction
for local starbursts, the implied bolometric surface-brightnesses
of the high-z galaxies are thus also very similar to local starbursts:
typically $\sim$ 10$^{10}$
to 10$^{11}$ L$_{\odot}$ kpc$^{-2}$. The high-redshift Lyman-break galaxies
appear to be `scaled-up' (larger and more luminous) versions of
local starbursts. The intrinsic UV surface brightnesses
are roughly three orders-of-magnitude higher than the disks of normal
spirals, and indeed normal spirals at $z \sim$ 3 (if they exist)
would be virtually undetectable
in $HST$ rest-UV images due to their low surface-
brightness (e.g. Hibbard \& Vacca 1997; Lanzetta \textit{et al.} 1999).

The similarity in the surface-brightnesses of local starbursts and Lyman-break
galaxies immediately implies that there are also strong similarities in
their basic physical properties. A high UV surface-brightness
implies a high star-formation rate per unit area ($\Sigma_{SFR}$)
and thus high surface-mass-densities in the stars ($\Sigma_*$) and the
interstellar gas ($\Sigma_g$)
that fuels the star-formation. A typical case would have
$\Sigma_{SFR} \sim$ 10 M$_{\odot}$ year$^{-1}$ kpc$^{-2}$ and
$\Sigma_g \sim  \Sigma_* \sim$ 10$^9$ M$_{\odot}$ kpc$^{-2}$.
These
are roughly 10$^3$ ($\Sigma_{SFR}$), 10$^2$ ($\Sigma_g$) and
10$^1$ ($\Sigma_*$) times larger than the
corresponding values in the disks of normal galaxies.

The basic physical and dynamical properties of starbursts and the Lyman-break
objects follow directly
from the above. A gas surface-mass-density of 10$^9$ M$_{\odot}$ kpc$^{-2}$
corresponds to an extinction of $A_B \sim$ 10$^2$ for a Milky Way
dust-to-gas ratio. The characteristic dynamical time in the star-forming region
will be short: $t_{dyn}$
$\sim$ $(G\rho)^{-1/2}$ $\sim$ $(G\Sigma_{tot}H)^{-1/2}$
$\sim$ few Myr, where $H \sim$ 10$^2$ pc is the thickness of the disk.
A surface-brightness of a few $\times$ 10$^{10}$ L$_{\odot}$ kpc$^{-2}$
corresponds to a radiant energy density inside the star-forming region that
is roughly 10$^3$ times the value in the ISM of the Milky Way. Finally,
simple considerations of hydrostatic equilibrium imply correspondingly
high total pressures in the ISM: $P \sim G \Sigma_g
\Sigma_{tot} \sim$ few $\times$ 10$^{-9}$ dyne cm$^{-2}$ (P/k $\sim$
few $\times$ 10$^7$ K cm$^{-3}$, or several thousand times the value
in the local ISM in the Milky Way). The rate of mechanical energy deposition
(supernova heating) per unit volume is also 10$^3$ or 10$^4$ times
higher than in the ISM of our Galaxy.

To summarize, the strong empirical similarity between local starbursts and the
high-z Lyman break galaxies implies strong similarity in their basic physical
properties. The conclusion seems inescapable:
{\it if we want to understand the Lyman break galaxies in the early
universe, we
need to understand local starbursts.}

As discussed by Heckman \textit{et al.} (1998), local
starbursts occupy a very small fractional volume in the multi-
dimensional manifold defined by such fundamental parameters as the
extinction, metallicity, and vacuum-UV line
strengths (both stellar and interstellar) of the starburst and the
rotation speed (mass) and absolute magnitude of the starburst's
`host' galaxy. In particular, more massive galaxies host more luminous,
more metal-rich, and dustier (more heavily-extincted) starbursts. There are 
simple physical
explanations for these trends. Firstly, simple considerations of causality
for a self-gravitating system with a gas mass $M_{gas} = f_{gas} M_{tot}$ 
imply that the maximum
possible star-formation rate is given by the conversion of all the gas into 
stars in one crossing time: $SFR_{max} \sim M_{gas}/t_{dyn}
\sim f_{gas} \sigma^3/G$. Thus, more massive galaxies (with larger velocity
dispersions $\sigma$) can sustain bursts with higher star-formation rates
and therefore larger luminosities. The physical basis for the strong 
observed correlation between galaxy mass and metallicity will be discussed
in \S3 below. Finally, the trend for more metal-rich starbursts to be more
heavily extincted
follows provided that neither the gas column density towards the starburst,
nor the fraction of interstellar metals locked into dust grains are
strong inverse functions of metallicity.

The result of these correlations is that an ultraviolet census of the local
universe would not only
underestimate the true star-formation-rate, it would systematically
under-represent the most powerful, most metal-rich starbursts
occurring in the most massive galaxies. This effect can be clearly seen
in the recent comparison of the vacuum-UV and far-IR galaxy luminosity functions
at low-redshift by Buat \& Burgarella (1998).
Estimates of star-formation rates at high-redshift based on
rest-frame vacuum-UV sample selection probably suffer the same bias, and
thus may under-
represent the ultra-luminous progenitors of the most massive present-day 
spheroids. It is
tempting to
speculate (based on the empirical properties of local starbursts) that
the $SCUBA$ far-IR-selected sources at high-z may represent just such objects.
This speculation is consistent with the relative space densities of
the most luminous Lyman-break galaxies and the $SCUBA$ sources (Meurer,
Heckman, \& Calzetti 1999).

{\it Clearly, we can not understand the
cosmic evolution of the star-formation rate and the formation and
evolution of galaxies without a panchromatic approach.}

\section{Starburst-Driven Superwinds}

Over the last few years, observations have provided convincing
evidence of the existence (and even the ubiquity) of `superwinds' -
galactic-scale outflows of gas driven by the collective effect of
multiple supernovae and stellar winds in a starburst (e.g. 
Lehnert \& Heckman 1996; Dahlem, Weaver, \& Heckman 1999; Martin 1999).

In this section I will summarize the systematic properties of superwinds 
as gleaned
from the analysis of their X-ray emission and the UV/optical absorption-lines.
The X-ray data have proven particularly important since they
are the only direct probe of the energetically-dominant `piston' of hot
gas that drives
the flow. The interstellar absorption-lines trace cooler
and denser gas that has probably been entrained into the outflowing hot gas
(e.g. Suchkov \textit{et al.} 1994), and
supply crucial
complementary data. They provide unambiguous information about the
magnitude and {\it sign} of the radial velocities of the gas, more fully
sample the whole range of gas densities in the
outflow (rather than being strongly weighted in favor of the densest material
which may contain relatively little mass), and can be
used to study outflows in high-redshift galaxies where the associated X-ray
or optical {\it emission} may be undetectably faint (Franx \textit{et al.}
1997; Pettini \textit{et al.} 1998).

\begin{figure}
\begin{center}
\begin{picture}(350,500)
\put(-30,-80){\includegraphics{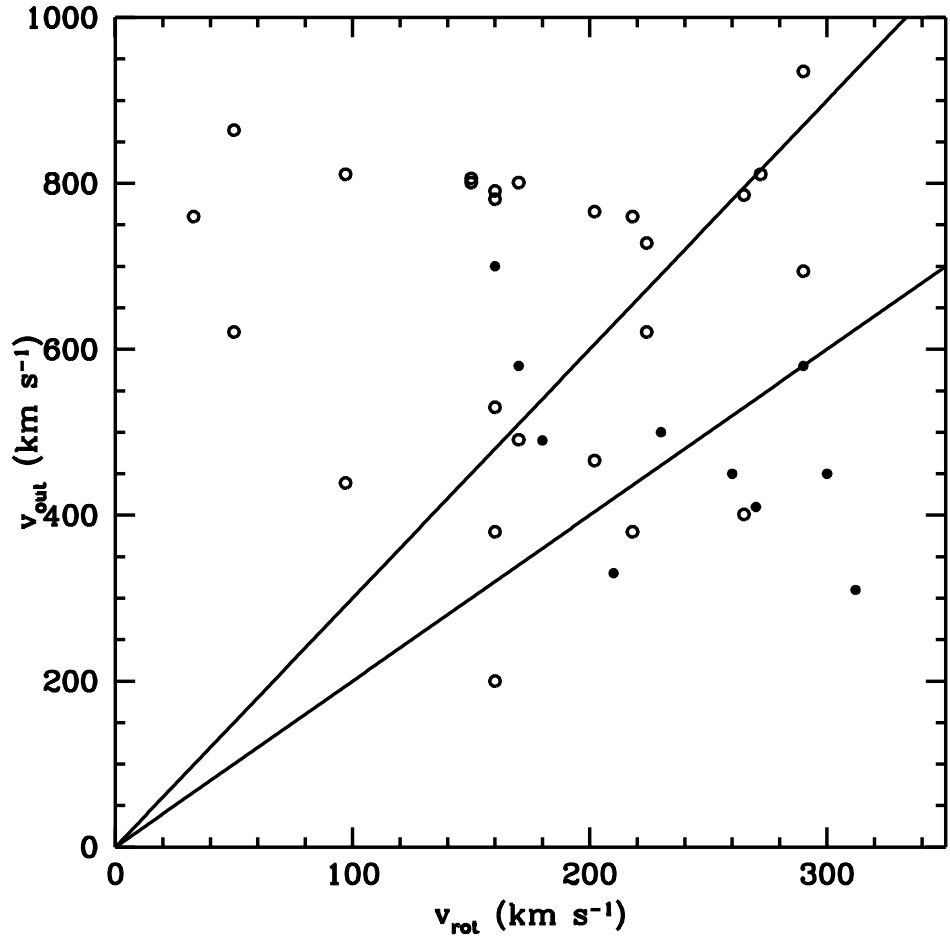}}
\end{picture}
\end{center}
\caption{Plot of the galaxy rotation speed
{\it vs.} the inferred terminal velocity of the outflow
for starburst galaxies. The two diagonal lines show the relations
$v_{out}$ = 2 $v_{rot}$ and $v_{out}$ = 3 $v_{rot}$. Data points
based on blueshifted interstellar absorption-line profiles are
indicated by solid dots and the points based on the X-ray temperatures
are indicated by hollow dots. Note that the two data sets are consistent with
each other,
imply that the outflow speed is independent of the host galaxy
potential well depth, and thus suggest that outflows will
preferentially escape from the least massive galaxies. See HLSA
for details.}
\end{figure}

Martin (1999) has empirically quantified
the global effects of intense star-formation on the 
interstellar medium by estimating the outflow rates
and velocities for a local sample of starburst and related galaxies.
I will follow her approach, but update it by adding new X-ray results on
additional galaxies, and improve upon it by including results from
our (Heckman \textit{et al.} 2000 - HLSA)  analysis of the interstellar
absorption-
lines in starbursts (a data type
not considered by Martin). The principal results are as follows:

\begin{itemize}
\item
{\it The outflow speeds are typically 400 to 800 km s$^{-1}$ independent
of the rotation speed of the `host' galaxy.}
This result is shown in figure 1. HLSA show that the absorbing gas typically
spans a velocity range from close to the galaxy systemic velocity up
to some maximum blueshift. They argued that this can be understood if
the hot outflow is ablating material off of cold dense clouds and
then accelerating it up to some terminal velocity. It is these inferred
terminal velocities that are plotted in figure 1. In the case of the
X-ray data, outflow velocities can not be directly observed. Instead,
we have estimated the outflow speed from the observed gas temperature
following Chevalier \& Clegg's (1985) solution for an adiabatic wind
fed by gas at a temperature T: $v \sim (5kT/\mu)^{1/2}$,
where $\mu$ is the mean mass per particle. This is a
conservative approach as it ignores any kinetic energy the
X-ray-emitting gas may already have. It is encouraging that the
overall agreement between the two data sets is satisfactory.

\item
{\it The implied outflow rates of metal-enriched material probably exceed 
the star-formation rate.}
I have calculated star-formation rates for a large sample of starburst
galaxies for which mass outflow rates can be estimated from either
interstellar absorption-lines, X-ray emission, or H$\alpha$
emission-line kinematics. The star formation rates assume
a Salpeter initial mass function extending from 0.3 to 100 $M_{\odot}$.
For the outflow rates based on
the absorption-line data, I follow HLSA and set these equal
to $\dot{M} \sim 60 (r_*/kpc) (N_H/3 \times 10^{21} cm^{-2}) (\Delta v/200 km/s)
(\Omega_w/4\pi)$ M$_{\odot}$/yr, where $r_*$ is the radius of the region
of mass-injection (the starburst), $N_H$ is the total column density
in the absorbing material, $\Delta v$ is the mean outflow velocity,
and $\Omega_w$ is the solid angle filled by the outflow. I adopted
the values measured or estimated for these parameters by HLSA,
Lehnert \& Heckman (1995), and Armus, Heckman, \& Miley (1990). For
the outflow rates derived from the X-ray data, I have simply taken the
inferred mass of hot gas (assuming a volume filling factor of unity)
and divided it by the outflow time ($\sim$ sound-crossing time) for the 
X-ray emitting region.
I also include the estimates for the outflow rates in the warm ionized
gas from Martin (1999), which are based on the observed velocities
in the H$\alpha$-emitting gas and the masses derived by assuming that
this gas is in pressure balance with the ram pressure of the superwind
(see Martin 1999 for details). While each method is crude, and makes
simplifying assumptions that may be unwarranted, it is gratifying
that the three methods yield similar results: the median value
for $\dot{M}/SFR$ =
1.4 for the absorption-line data, 2.8 for the X-ray data, and
1.6 for the H$\alpha$ data. 
In fact, since these three
methods measure different gas phases, the {\it total} outflow rate
could be approximated as the sum of the three rates. Thus,
gas is being
expelled from starbursts at a rate in excess of the rate that gas is
being processed into stars.

\item
{\it The implied outflow rates of kinetic/thermal energy are a significant
fraction of the total injection rate by supernovae and stellar winds.}
The rate at which superwinds carry energy is uncertain. Simple estimates
can be made from the X-ray data by calculating the thermal energy content
in the hot gas (assuming unit volume filling factor) and dividing this
by the dynamical time for the X-ray-emitting region. As emphasized by
Strickland (1998), this ignores the
effects of clumping (which will reduce the thermal energy) and of the
supersonic flow speeds (which represent significant kinetic energy). 
Energy (more
properly momentum) outflow rates can also be deduced from the radial
pressure gradients observed via standard optical emission-line diagnostics,
assuming that these trace the sum of the ram- and thermal pressure in
the outflow (e.g. Heckman, Armus, \& Miley 1990; Lehnert \& Heckman
1996). Both techniques suggest that the majority of the kinetic/thermal
energy supplied by the supernovae and stellar winds is being carried
out in the flow (and has not been radiated away). The key seems to
be that the porosity of the starburst ISM is high, so that most supernovae
detonate in a rarefied pre-heated medium (e.g. Heckman, Armus, \& Miley 1990;
Marlowe \textit{et al.} 1995).

\item
{\it The outflows are apparently quite dusty, with inferred reddenings of
$E(B-V) \sim$ 1 over regions of a-few-to-ten kpc in size.}
HLSA mapped the Na-$D$ interstellar
absorption-line over regions with sizes of a few
to ten kpc in a sample of 18 starbursts.
Figure 2 shows the strong correlation they found between the 
depth of the absorption-line profile (roughly proportional
to the fraction of the emitting region covered by absorbing gas)
and the line-of-sight reddening towards the emitting region. The
observed reddening is substantial, with $E(B-V) \sim$ 0.9$\pm$0.4
(corresponding to $N_H \sim$ several $\times$ 10$^{21}$
cm$^{-2}$ for normal Galactic dust). This generalizes Phillips's (1993) 
discovery of a spectacularly dusty few-kpc-scale outflow in the halo of
the starburst galaxy NGC1808.
\end{itemize}

\begin{figure}
\begin{center}
\begin{picture}(350,500)
\put(-30,-80){\includegraphics{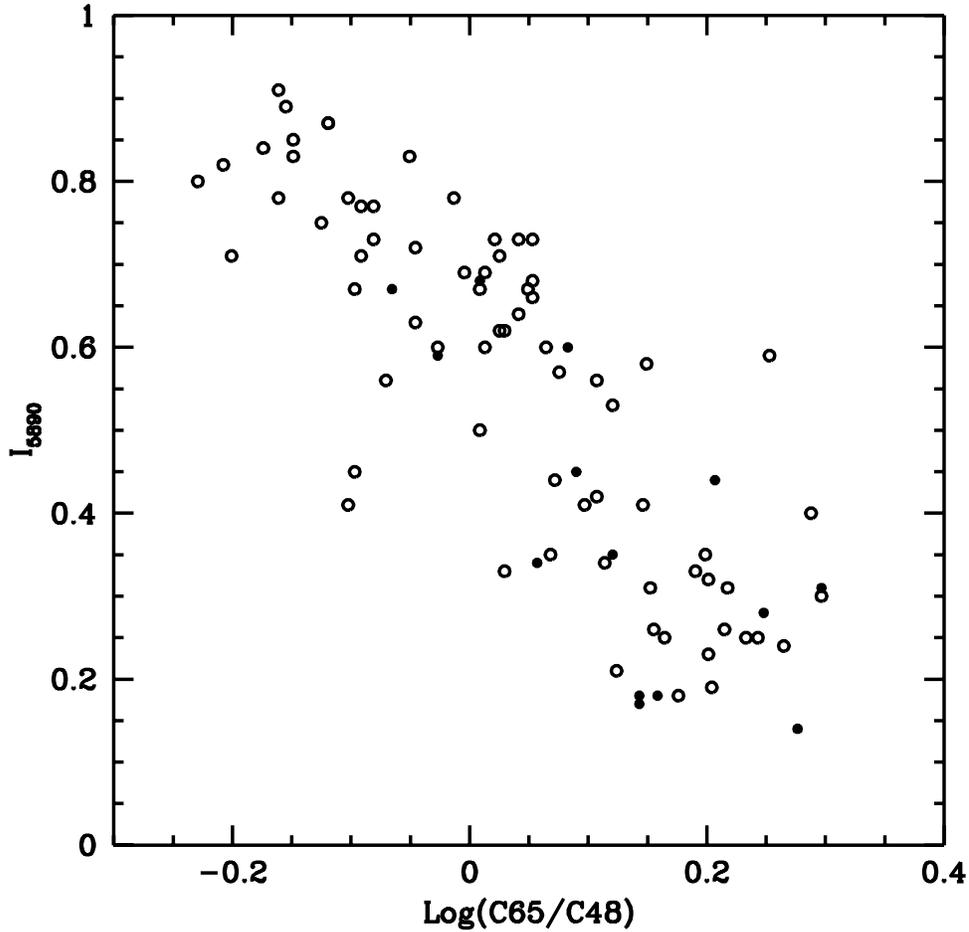}}
\end{picture}
\end{center}
\caption{Plot of the normalized residual intensity at the
center of the interstellar NaI$\lambda$5890
transition ($I_{5890}$) {\it vs.} the log of the color of the optical continuum
(the ratio of $F_{\lambda}$ at rest wavelengths of 6560 and 4860
\AA). Points plotted as solid dots are the nuclei of powerful starbursts.
Other points
are off-nuclear locations.
The deeper the
NaI absorption-line, the more-reddened the background starlight.
An unreddened starburst population should have $log(C_{65}/C_{48})$
= -0.3. For a standard Galactic reddening curve, the implied
$A_V$ ranges up to roughly 4 magnitudes for the most-reddened sight-lines.
See HLSA
for details.}
\end{figure}

The three results above have a variety of important implications.
The independence of the outflow speed on the galaxy rotation speed
strongly suggests that the outflows selectively escape the potential
wells of the less massive galaxies, carrying metals with them.
This process plausibly accounts for
the strong mass-metallicity relation in ellipticals and bulges.
Lynden-Bell (1992) has proposed an appealingly simple model in which
the fraction
of starburst-produced metals
that are retained by a galaxy experiencing an outflow is proportional to the
galaxy potential-well
depth for galaxies with $v_{esc} < v_{out}$, and asymtotes to
full retention for the most massive galaxies ($v_{esc} > v_{out}$).
For $v_{out}$ in the range we measure (figure 1), such a simple prescription
can reproduce the observed mass-metallicity relation for spheroids
(Lynden-Bell 1992).

At the same time, superwinds could
deposit the required amount of observed metals in the
intra-cluster medium (e.g. Gibson, Loewenstein, \& Mushotzky 1997).
The kinetic energy carried out by superwinds at high redshift could
be the agent that `preheated' the proto-intracluster medium prior
to the collapse/formation of rich clusters (Ponman, Cannon, \&
Navarro 1999; Evrard, this volume).
If the ratio of ejected metals to stellar
spheroid mass is the same globally as in clusters of galaxies,
the present-day
mass-weighted  metallicity of a general intergalactic medium
with $\Omega_{igm}$ = 0.015 will be $\sim$ 1/6 solar (see also
Renzini 1997). These intergalactic metals could reconcile the apparent
deficit in the inventory of metals contained by stars in the
present universe compared to the expectations based on the integrated
amount of high-mass star-formation over cosmic time (e.g.
Aguirre 1999). Pettini (this volume) notes a similar problem with
`missing metals' at high-z. Highly-ionized superwind ejecta may
be a plausible place to hide them. Superwinds may also help alleviate
the problems with understanding the apparently wide distribution
of metals in the Ly-$\alpha$ forest (Pettini, this volume;
Efstathiou, this volume).

More speculatively, {\it if} the outflowing dust survives its journey,
the cumulative effect of dusty superwinds
could lead to a cosmologically-significant
amount of intergalactic dust 
($\Omega_{dust}$ = few $\times$ 10$^{-5}$), which would seriously
affect the Hubble Diagram for type Ia supernovae (Aguirre 1999;
Aguirre \& Haiman 1999). Finally, Kronberg, Lesch, \& Hopp (1999) have
argued that a substantial fraction of the inter-galactic medium at
high-redshift
can be permeated with magnetic fields carried out of
the first generation of galaxies by early superwinds.

\section{The Starburst-Seyfert Connection}

There have been many observational investigations into the possible
presence of starbursts in Seyfert nuclei, and it is
beyond the scope of this contribution to review this literature. Instead,
I want to briefly report on efforts by our group to make direct spectroscopic
detections of hot, massive stars in a large, unbiased sample of Seyfert nuclei.

According to the standard `unified' picture for radio-quiet AGN
the principal building blocks for a Seyfert nucleus
are: 1. A supermassive black hole and its associated accretion disk
(the primary source of X-ray through optical continuum emission) 
2. An optically-
and geometrically-thick circum-nuclear torus of dust and gas, with an inner
radius of a few pc and an ill-defined outer radius (of-order 10$^2$ pc).
It is viewed close to its equatorial plane (polar axis) in type 2
(type 1) Seyferts.
3. A `mirror' of dust and/or warm electrons located along the polar axis
of the torus. Thus, the optimal targets in which to search for hot stars
are the type 2 Seyfert nuclei, in which the torus
providentially blocks out the
blinding glare from a hidden type 1 Seyfert nucleus. Indeed,
type 2 Seyfert nuclei have long been known to exhibit a
`featureless continuum' (`$FC$') that produces most of the UV light and
typically 10\% to 30\% of the visible/NIR light (the rest appears
to be light from an ordinary old population of stars). Until
recently, it was thought that the optical/UV $FC$ was mostly light from the
hidden type 1 Seyfert nucleus that had been reflected into our
line-of-sight by the mirror.

Instead, if at least part of the $FC$ is produced by a population of
hot stars, it should not
actually be featureless! Instead, spectroscopy in the blue and near-UV
should show the high-order Balmer lines in absorption (unlike
H$\alpha$ and H$\beta$, which are dominated by nebular emission)
and weak HeI stellar photospheric lines (Gonzalez-Delgado 
\textit{et al.} 1999).
Spectra in the vacuum-UV should reveal strong stellar wind lines and 
weaker stellar photospheric lines (cf. de Mello, Leitherer, \& Heckman
1999).
To test this possibility, we
have therefore undertaken a program to obtain high-resolution
vacuum-UV images and spectra (with
$HST$) and near-UV spectra (with ground-based telescopes) of a
representative sample of the brightest type 2 Seyfert nuclei.

The first results have been presented in detail in Heckman
\textit{et al.} (1997)
and Gonzalez-Delgado \textit{et al.} (1998).
$HST$ imaging shows
that the UV continuum source in every case is spatially-resolved
(scale size few hundred pc or greater). In some cases the
morphology is strikingly reminiscent of UV images of starbursts.
In other cases,
a component of the UV continuum is
roughly aligned with the inferred polar axis of the obscuring torus
(as expected for reflected and/or reprocessed light from the
central engine).

Of the original sample of 13 type 2 Seyfert with $HST$ vacuum-UV images, only
four were bright enough for us to obtain spectra of adequate quality
in the crucial UV spectral window from about 1200 to 1600~\AA.
However, these spectra are decisive: all four show the clear
spectroscopic signature of a starburst population that dominates
the UV continuum. In addition to classic strong stellar
wind features (NV$\lambda$1240, SiIV$\lambda$1400, and
CIV$\lambda$1550), we can also detect weaker and much narrower
absorption features from highly-excited transitions (which are
therefore indisputably of stellar origin).

In each of the four cases, if we use the empirical `starburst
attenuation law' (Calzetti, Kinney, \& Storchi-Bergmann 1994;
Meurer, Heckman, \& Calzetti 1999) to
correct the observed UV continuum for dust extinction, we find that
the bolometric luminosity of the nuclear (10$^{2}$-pc-scale)
starburst is comparable to the estimated bolometric luminosity
of the `hidden' type 1 Seyfert nucleus (of-order 10$^{10}$
L$_{\odot}$). Large-aperture UV spectra with $IUE$ imply
the existence of a surrounding larger-scale (few kpc)
and more powerful (few $\times$ 10$^{10}$ to
10$^{11}$ L$_{\odot}$) dusty starburst that is energetically capable
of powering the bulk of the observed far-IR emission from the galaxy.
Thus, starbursts are an energetically
significant (or even dominant) component of at least {\it some}
Seyfert galaxies.

However, we have $HST$ spectra of only four type 2 Seyferts, and
these are strongly biased in favor of cases with high UV surface-
brightness. Can we say anything more general? To address this, we
have embarked on a program to obtain spectra from about 3500 to
9000~\AA\ of a complete sample of the 25 brightest type 2 Seyfert
nuclei in the local universe. These objects are selected from
extensive lists of known Seyfert galaxies on the basis of the flux
of either the nebular line-emission from the Narrow Line Region
(the [OIII]$\lambda$5007 line) or of the nuclear radio source
(Whittle 1992).

We are still analyzing these spectra, but even a cursory inspection
of the near UV region (below 4000~\AA) shows that about half have
pronounced Balmer absorption-lines whose strength is consistent
with a population of late O or early B stars. In several cases we
can also detect those photospheric HeI absorption-lines ($\lambda$4921,
$\lambda$4387, $\lambda$3819) that are not filled in by nebular emission.
In most of
the remainder of the sample, the $FC$ is so weak relative
to the light from a normal old-bulge stellar population that its
origin is still not clear. 
Recent work in a related vein has been undertaken by Cid Fernandes,
Storchi-Bergmann, \& Schmitt (1998)
and Schmitt, Storchi-Bergmann, \& Cid Fernandes (1999). Their
results agree at least qualitatively with ours:
they find that most of the optical and near-UV $FC$ in type 2 Seyfert
nuclei is produced by young and intermediate age stars
(age $\leq$ 100 Myr).

Thus, it is clear that massive stars and starbursts play an important
energetic role in a significant fraction of Seyfert nuclei. What is
not yet clear is whether starbursts are an {\it essential}
component of the Seyfert phenomenon and what (if any) the causal
or evolutionary connection might be. Perhaps - as Cid Fernandes
\& Terlevich (1995) suggested - the starburst is an inevitable
byproduct of the dense molecular torus that is now believed
to be a fundamental part of the inner machinery of AGN
(both obscuring the `central engine' and serving as its fuel source).
If true, this would have major implications for
the relationship between quasars and galaxy formation.

\section{The Host Galaxies of High-z Radio-Quiet Quasars}

The cosmic evolution of the population of powerful radio galaxies has
been well-documented over the past decade (see the volume edited by 
Rottgering, Best,
\& Lehnert 1999). The uniformity of the K-band Hubble Diagram out
to $z \sim 4$ and
the evolution of the red-envelope in the visible and near-IR colors
are consistent with a large redshift of formation and the subsequent
passive evolution of the progenitors of (some) present-day giant elliptical
galaxies (e.g. Lilly 1989; Eales \& Rawlings 1996; McCarthy 1999).
Much less is known about the evolution
of the population of the hosts of radio-loud quasars, but the available data 
paint a broadly similar picture (e.g. Lehnert \textit{et al.} 1992;
Ridgway \& Stockton 1997; McLure \textit{et al.} 1999). This is
consisent with the standard `unified model'
in which radio-loud quasars and radio galaxies are drawn from the same
parent population, but the quasars (radio galaxies) are viewed
roughly along (perpendicular to) the polar axis of a dusty torus
(Barthel 1989).

While the notion of the ``monolithic'' formation of apparently-massive
elliptical
galaxies at high redshifts does not fit comfortably into the standard
CDM picture of hierarchical assembly at late times, powerful radio
galaxies are exceedingly rare objects. For H$_0$ = 50 km s$^{-1}$
Mpc$^{-1}$ and $\Omega$ = 1,
3CR radio galaxies at z = 2.5
have a co-moving space density of only 0.2 Gpc$^{-3} \Delta$log$P_{rad}^{-1}$
while the fainter 6C (Eales 1999) and MRC (McCarthy 1999) radio
galaxies have space densities roughly $10^2$ times larger (Dunlop
\& Peacock 1990).
These values
can be compared to the present-day space-density of first-ranked cluster
galaxies (roughly 5000 Gpc$^{-3}$ - Bahcall \& Cen 1993). Even allowing for
a short lifetime
for the radio galaxy phase, 
the evolved descendants of {\it powerful} radio galaxies would account for
only a small minority of the first-ranked cluster elliptical galaxies.

In contrast, radio-quiet quasars are far more common, and therefore more
likely to be progenitors of typical present-day early-type galaxies.
Radio-quiet quasars with $M_B \leq$ -23 have co-moving space densities
at z$\sim$2 of $\sim$2$\times$ 10$^4$ Gpc$^{-3}$
(Hartwick \& Schade 1990). Now, for a quasar lifetime of-order the
Eddington growth-time 
(a few \% of the 
Hubble time at z = 2), the implied space density of the present-day
descendants is of-order 10$^6$ Gpc$^{-3}$, which is comparable to the space
density of $L_*$ E's and S0's (e.g. Fukugita, Hogan, \&
Peebles 1998). This identification is quite consistent with
the correlation between the masses of supermassive black hole and
the spheroids in which they live today. A quasar with $M_B$ = -24 
powered by accretion at the Eddington rate requires $M_{SMBH} \sim$
2 $\times$ 10$^8$ $M_{\odot}$, and this supermassive black hole
would live today in spheroid with a mass of $M_{sph,z=0} \sim$ 5 $\times$
10$^{10}$ $M_{\odot}$ and V-band luminosity of about 1.5 $\times$
10$^{10}$ $L_{\odot}$ $\sim$ 0.4 $L_*$
(Magorrian \textit{et al.} 1998;
van der Marel 1999).

Thus, it is clearly important to document the properties of the host
galaxies of radio-quiet quasars over a broad range in redshifts.
At low redshifts, the hosts of the {\it most-luminous}
radio-loud quasars, radio-quiet 
quasars, and radio galaxies all appear to be similar: several-L$_*$ 
E or S0 galaxies (McLure \textit{et al.} 1999). However,
ground-based near-IR imaging of small samples already hinted that
the situation might be quite different at high-redshift, with the
hosts of radio-quiet quasars being significantly fainter than their
radio-loud cousins (Lowenthal \textit{et al.} 1995; but see
Aretxaga \textit{et al.} 1998).

Recent analyses of $HST$ NICMOS images of radio-quiet quasars at
$z \sim 2$ have now clarified the situation. We (Ridgway 
\textit{et al.} 1999 - RHCL) have imaged 5 faint (M$_B \sim$ -23) radio-quiet
quasars, complementing the analysis reported by Rix \textit{et al.} 1999)
of six luminous ($M_B \sim$ -26) gravitationally-lensed cases. Some
examples of the underlying hosts galaxies in the RHCL sample are shown
in figure 3.

\begin{figure}
\begin{center}
\begin{picture}(350,500)
\put(-30,-80){\includegraphics{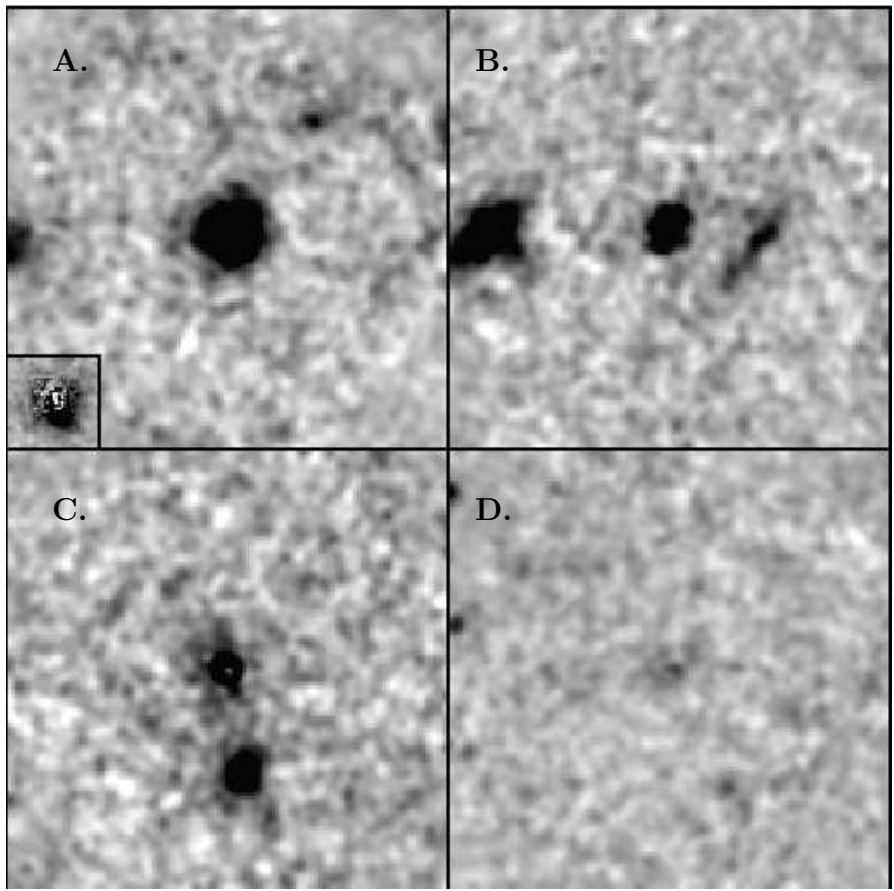}}
\put(20,320){\large{\bf A.}}
\put(180,320){\large{\bf B.}}
\put(20,150){\large{\bf C.}}
\put(180,150){\large{\bf D.}}
\end{picture}
\end{center}
\caption{High-z radio-quiet quasar hosts, after the quasar has been subtracted
and the image smoothed with a Gaussian
kernel of 0.06 arcsec. Each panel is 5.7 arcsec square
(or roughly 45 kpc), N up, E left.
{\it A.} MZZ 9592, $z \sim $ 2.7,
with an inset of the central region (unsmoothed).
There is an off-center residual host component.
{\it B.} MZZ 9744, $z \sim$ 1.8
{\it C.} MZZ 1558, $ z \sim$ 2.7 {\it D.} MZZ 4935, $z \sim $1.8,
host marginally detected. Note the apparent close ($\sim$ 10 kpc) companion
galaxies in panels {\it B} and {\it C}.
See RHCL for details.}
\end{figure}

\begin{figure}
\begin{center}
\begin{picture}(350,500)
\put(-10,-30){\includegraphics{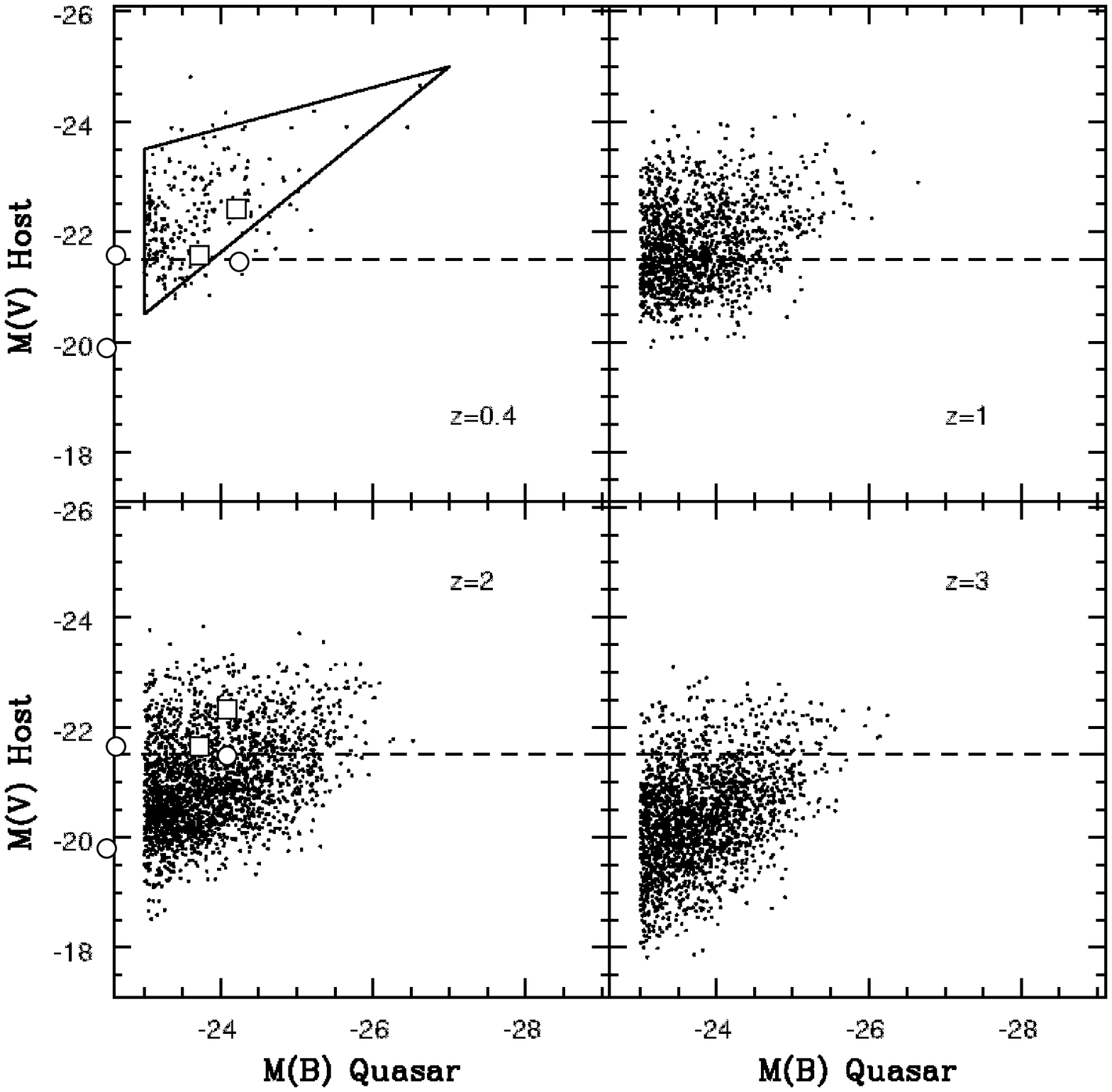}}
\end{picture}
\end{center}
\caption{The measured properties of the five radio-quiet $z \sim$ 2 quasars
and hosts in the RHCL sample (hollow symbols) overplotted on the theoretical
predictions from Kauffmann \& Haehnelt (1999 - their figure
12). The data have been plotted at both $z$ = 2 (bottom left panel)
to compare to the models (small solid points) and at $z$ = 0.4
(top left panel) to compare to observations of low-z hosts (
indicated by the triangular region). See RHCL for details.
Rix \textit{et al.} (1999) find similar results for more luminous
$z \sim$ 2 radio-quiet quasars.}
\end{figure}

While the samples are still modest in size, several conclusions can already
be drawn (see figure 4; RHCL; Rix \textit{et al.} 1999):

\begin{itemize}

\item
Typical radio-quiet quasars at $z \sim$ 2 are hosted by galaxies with
rest-frame absolute visual magnitudes similar to present-day
$L_*$ galaxies ($M_V$ = -20 to -23 = $M_{*,V}\pm$1.5 mag).

\item
As such, they are much fainter than radio galaxies at the same redshift
(typically by $\sim$ 2 magnitudes).

\item
These host galaxies are -if anything- {\it less} luminous than the
hosts of similarly-powerful low-z radio-quiet quasars.
Since the luminosity-weighted mean age of the stellar population
in the high-z hosts is almost certainly younger than that of
the low-z hosts, the difference in stellar {\it mass} will be even more
pronounced.

\item
The rest-frame-visual luminosities and sizes of the radio-quiet
quasar hosts are roughly similar to those of the Lyman-break galaxies.
Thus, the Lyman-break population {\it might} represent the parent
population of typical radio-quiet quasars. Our Cycle 8 HST
program will determine whether this similarity extends into
the rest-frame UV.

\end{itemize}

The potential implications of these results are quite tantalizing.
First, they imply that the well-studied cosmic evolution of the 
hosts of the very-radio-loud AGN population 
is evidently {\it not} representative of the much more numerous radio-quiet
population. Second, if we make the simplifying assumptions
that the ratio of $L_{quasar}/M_{SMBH}$ is roughly independent of
redshift and that $M_{SMBH} \propto M_{sph,z=0}$, then it follows that
supermassive black holes form well before their host galaxies are fully
assembled.
This agrees qualitatively with the idea of the hierarchical assembly
of massive galaxies at late epochs. Indeed, as already pointed out
by Rix \textit{et al.} (1999) and RHCL, the 
observations (figure 4) agree rather well
with the recent theoretical predictions of Kauffmann \& Haehnelt
(1999).

\section{Summary}

If the reader takes away only four ideas from my contribution,
I hope they are the following:\\

1. Starburst galaxies are good analogues (in fact, the only plausible
local analogues) to the known population of star-forming galaxies at
high-redshift.\\

2. Integrated over cosmic time, supernova-driven galactic-winds
(`superwinds') play an essential role
in the evolution of galaxies and the inter-galactic medium.\\

3. Circumnuclear starbursts are an energetically-significant
component of the Seyfert phenomenon.\\

4. The evolution of the population of the host galaxies of radio-quiet
quasars is significantly different than that of powerful radio galaxies,
and is at least qualitatively consistent with the standard picture
of the hierarchical assembly of massive galaxies at relatively late
times.\\[5mm]

\newpage
{\bf Acknowlegments}

I would like to thank all my collaborators on the work described above,
and in particular Rosa Gonzalez-Delgado, and Susan Ridgway. My research
is supported in part by NASA LTSA grant NAG5-6400 and HST
grant GO-07864.

\end{document}